%% file: main.tex
\documentclass[journal]{IEEEtran}
\usepackage{titlesec}
\titlespacing*{\section}{0pt}{*1}{*1}
\titlespacing*{\subsection}{0pt}{*0.5}{*0.5}
\usepackage{multirow}
\usepackage{mathtools}
\usepackage{amsmath,amssymb,amsfonts}
\usepackage{graphicx}
\usepackage{textcomp}
\usepackage{xcolor}
\usepackage{subfigure} 
\usepackage[letterpaper, left=0.65in, right=0.65in, bottom=1in, top=0.70in]{geometry}

\def\BibTeX{{\rm B\kern-.05em{\sc i\kern-.025em b}\kern-.08em
    T\kern-.1667em\lower.7ex\hbox{E}\kern-.125emX}}
\usepackage{xcolor}
\usepackage[linesnumbered,ruled,vlined]{algorithm2e}
\usepackage{listings}
\usepackage[hidelinks,colorlinks=true,linkcolor=blue,citecolor=blue,urlcolor=black]{hyperref}
\usepackage[noend]{algpseudocode}

\SetCommentSty{mycommfont}
\SetKwInput{KwInput}{Input}   
\SetKwInput{KwOutput}{Output} 

\definecolor{GreenForest}{rgb}{0.09, 0.45, 0.27}
\usepackage{soul}
\usepackage{balance}
\usepackage{cite}
\usepackage{acro} 
\input{acro.tex}

\usepackage{comment}
\usepackage{amsmath}
\usepackage{steinmetz}
\makeatletter
\newcommand{\ie}{\emph{i.e.}\@ifnextchar.{\!\@gobble}{}}
\newcommand{\eg}{\emph{e.g.}\@ifnextchar.{\!\@gobble}{}}
\newcommand{\etc}{etc\@ifnextchar.{}{.\@}}
\makeatother

\usepackage{enumitem}

\begin{document}

\bstctlcite{IEEEexample:BSTcontrol}
    \title{Single-Carrier Waveform Design for Joint Sensing
and Communication}
  \author{Ayoub Ammar Boudjelal, Rania Yasmine Bir, and H\"{u}seyin Arslan,~\IEEEmembership{Fellow,~IEEE}

\thanks{The authors are with the Department of Electrical and Electronics Engineering, Istanbul Medipol University, Istanbul, 34810, Turkey (e-mail: 
ayoub.ammar@std.medipol.edu.tr; rania.bir@std.medipol.edu.tr; huseyinarslan@medipol.edu.tr).}
}

\IEEEpeerreviewmaketitle
\maketitle

\begin{abstract}
The emergence of 6G wireless networks demands solutions that seamlessly integrate communication and sensing. This letter proposes a novel waveform design for joint sensing and communication (JSAC) systems, combining single-carrier interleaved frequency division multiplexing (SC-IFDM), a 5G communication candidate signal, with frequency modulated continuous wave (FMCW), widely used for sensing. The proposed waveform leverages the sparse nature of FMCW within SC-IFDM to achieve orthogonal integration in three steps: SC-IFDM symbols are allocated alongside the sparse FMCW, followed by the SC-IFDM transform into the time domain, and a cyclic prefix (CP) is applied in which phase shifts are introduced to the FMCW. Additionally, an enhanced channel estimation method is incorporated to boost system performance. Simulation results demonstrate the proposed waveform's ability to deliver high-resolution sensing and superior communication performance, surpassing traditional multicarrier designs.

\end{abstract}

\section{Introduction}

\IEEEPARstart{T}{he} \ac{JSAC} paradigm is increasingly recognized as essential for deploying efficient \ac{6G} networks. By integrating communication and sensing functionalities, \ac{JSAC} enables network nodes to not only transmit data but also perceive their environment, leading to improved system reliability, decision-making, and overall resource efficiency \cite{luong2021radio,nguyen2022access}. However, as device density increases, interference becomes a pressing issue, highlighting the need for novel sensing and communication waveform designs to manage resources effectively while ensuring interference mitigation and supporting the stringent requirements of real-time sensing and communication in \ac{6G} \cite{luong2021radio}.

Several signal designs have been presented to perform both sensing and communication, such as \ac{OFDM} and \ac{OTFS} \cite{gaudio2020effectiveness}. These multicarrier signals demonstrate estimation accuracy comparable to that of \ac{FMCW}, a widely used automotive radar waveform \cite{patole2017automotive,zhou2022integrated}, while simultaneously achieving high communication rates. These findings indicate that joint radar estimation and communication can be efficiently performed without compromising throughput or accuracy.

In \cite{zegrar2024novel}\cite{zegrar2022otfs}, \ac{OTFS} and \ac{FMCW} signals are combined to achieve high data rates and low-complexity sensing. However, \ac{OTFS} suffers from high \ac{PAPR}, reducing power efficiency and requiring pre-coding techniques \cite{xiong2024expanded}. While the coexistence aims to simplify sensing and increase throughput, it introduces complexity by relying on radar processing for delay and Doppler shifts, undermining \ac{OTFS}'s benefits. In \cite{bouziane2024novel}, the coexistence of \ac{OFDM} and \ac{FMCW} achieves comparable sensing performance, but reallocating diagonal elements disrupts orthogonality, degrading communication performance. Additionally, the communication performance was only tested in static channels, making it impractical for mobile environments.\\
 In single-carrier systems, \ac{SC-IFDM}, interpreted as a linearly precoded \ac{OFDM} in which the occupied sub-carriers are equally spaced over the entire bandwidth \cite{fang2012interleaved}, offers a key advantage over multicarrier signals where lower \ac{PAPR} level can be achieved. This low \ac{PAPR} level simplifies the transmitter design for user devices and makes it ideal for uplink transmission
in LTE systems \cite{myung2006peak,sahin2016flexible} and \ac{6G} applications.

Building on the strengths of \ac{SC-IFDM}, this letter introduces a novel waveform design for \ac{JSAC} systems by integrating \ac{FMCW} signal orthogonally with \ac{SC-IFDM} signaling. The key contributions of this work are: \footnote{\textit{Notation:} Bold uppercase $\mathbf{A}$, bold lowercase $\mathbf{a}$, and unbold letters $A,a$ denote matrices, column vectors, and scalar values, respectively. $(\cdot)^{-1}$ denote the inverse operators.  $\delta(\cdot)$ denotes the Dirac-delta function.  $\mathbb{C}^{{M\times N}}$ denotes the space of $M\times N$ complex-valued matrices.}
\begin{itemize} [leftmargin=*]
    \item A novel \ac{SC-IFDM}-\ac{FMCW} signal is proposed, enabling simultaneous communication and sensing. This design leverages the sparsity of \ac{FMCW} in the \ac{DFT} domain to diagonally integrate it with the \ac{SC-IFDM} signal, minimizing interference and simplifying the receiver architecture for both functions.
    \item A novel channel estimation method is proposed, using the integrated FMCW signal as a pilot in the DFT domain. This eliminates the need for guard bands between pilot and data symbols, leveraging the chirp's sparsity in the DFT domain rather than the frequency domain, thereby boosting spectral efficiency.
    \item Simulation results demonstrate that the proposed waveform achieves high-resolution sensing and outperforms JSAC waveforms like OFDM-FMCW and OTFS-FMCW in communication performance.

\end{itemize}

\section{THE PROPOSED WAVEFORM DESIGN}
\begin{figure}[t]
   \centering
\includegraphics[width=0.47\textwidth]{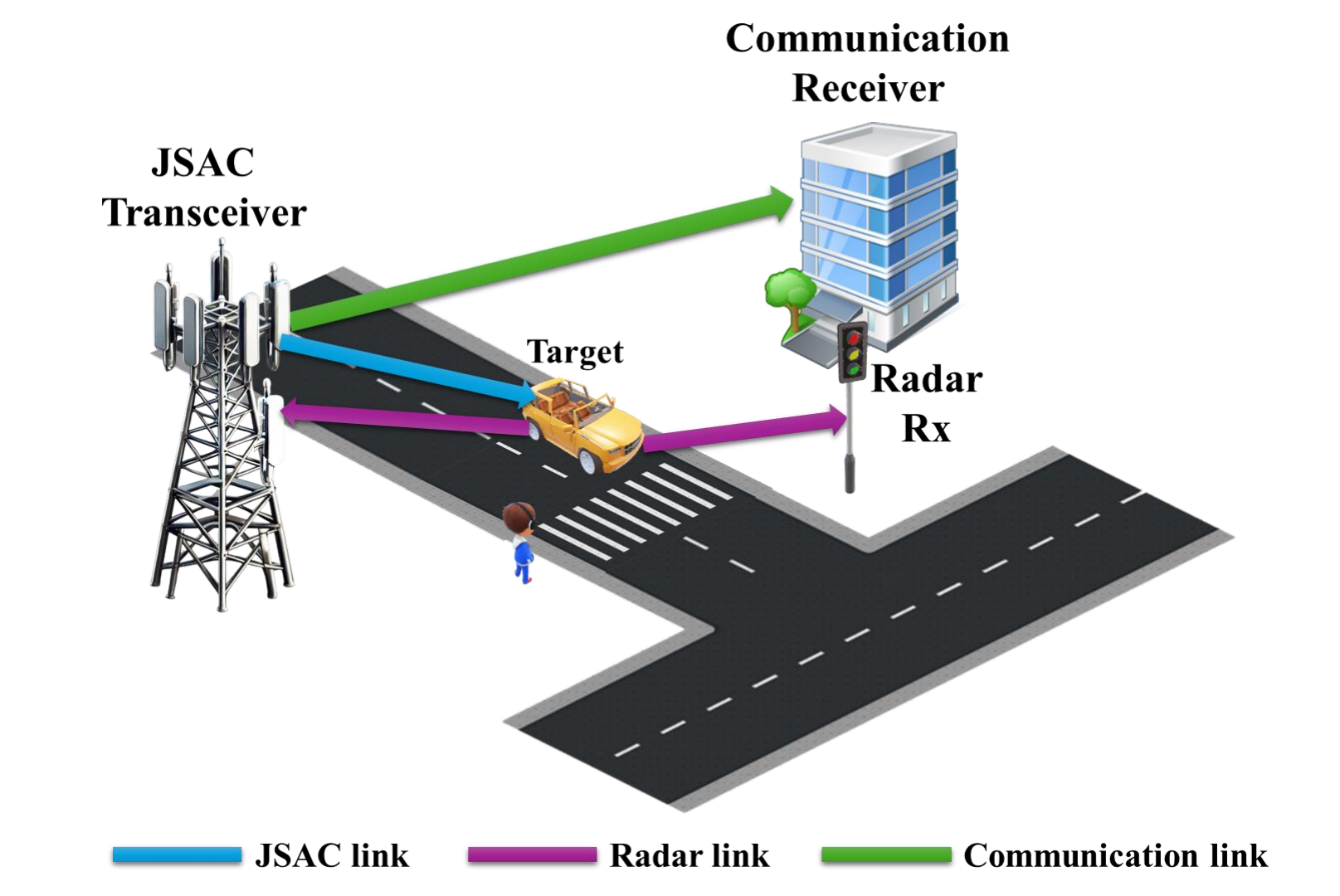}
    \caption{Illustration of the JSAC network.}
    \label{fig:JSAC_system}
\end{figure}

\hspace{1em}In this section, we model the \ac{SC-IFDM} and \ac{FMCW} signaling and describe how the sparse representation of \ac{FMCW} integrates with \ac{SC-IFDM} to create the proposed \ac{SC-IFDM}-\ac{FMCW} waveform for \ac{JSAC} systems. The proposed \ac{JSAC} system performs dual functions, serving as a mono-static radar for sensing and a communication transmitter. Additionally, other separate systems in the environment may act as bi-static radars for sensing, as illustrated in Fig.\ref{fig:JSAC_system}. By transmitting the proposed \ac{SC-IFDM}-\ac{FMCW} waveform, the system simultaneously detects radar echoes for sensing while transmitting data to a dedicated receiver, ensuring there is no interference between the sensing and communication functionalities.

\subsection{SC-IFDM Signal Model}
Assuming \( M \times N \) data symbols are modulated using \ac{SC-IFDM} by mapping them to \( N \) \ac{DFT} blocks, each of size \( M \), and interleaving the output of these DFT blocks before applying the final \( MN \)-point \ac{IDFT}, the resulting signal \( s^{\text{SC-IFDM}}(p) \), with \( \{X^{\text{SC-IFDM}}(k,l)\} \) for \( k=0, \dots, N-1 \) and \( l = 0, \dots, M-1 \), is expressed as \cite{chia2006distributed}
\vspace{-5pt}
\begin{equation}
    s^{\text{SC-IFDM}}(p)=\frac{1}{\sqrt{N}} \sum_{k=0}^{N-1}  X^{\text{SC-IFDM}}(k,[p]_M)e^{j 2 \pi\frac{k}{MN}p},
    \label{equ:simplify11}
\end{equation}
for $p=0,1,\dots,MN-1$.
We can reformulate \eqref{equ:simplify11} by setting $ p=l+n M$ , such that $ n=0,\dots,N-1$.  Substituting this into the expression, we get
\begin{equation}
    s^{\text{SC-IFDM}}(l+n M)=\frac{1}{\sqrt{N}} \sum_{k=0}^{N-1} X^{\text{SC-IFDM}}(k,l) e^{j 2 \pi\frac{k(l+nM)}{MN}}.
    \label{equ:simplifypp}
\end{equation}

\subsection{FMCW Signal Model}
\hspace{1em}Consider a linear \ac{FMCW} signal, $s^{\text{FMCW}}(t)$, which sweeps linearly across bandwidth $B_c$ over a time duration $T_c$. The time-domain representation of $s^{\text{FMCW}}(t)$ is expressed as \cite{csahin2020multi} 
\begin{equation}
    s^{\text{FMCW}}(t)=e^{j\pi  \eta t^2 }, \quad 0 \leq t<T_c,
\label{equ:chirp}
\end{equation}
where $\eta=B_c / T_c$ is the \ac{FMCW} rate. Here, $\eta$, $T_c$, and $B_c$ $\in$ $\mathbb{R}$. 
Assume that the time-bandwidth product $MN=\eta T_c^2$ is a nonzero integer. The critically sampled FMCW pulse is then formed by sampling at discrete points $t=\frac{p}{MN}T_c$, resulting in
\begin{equation}
s^{\text{FMCW}}(p)=e^{j\pi  \frac{ p^2 }{MN} } , \quad 0 \leq p<MN.
\label{equ:discrete_chirp}
\end{equation}
\subsection{SC-IFDM to FMCW}
\hspace{1em}Taking $p=l+n M$, the discrete \ac{FMCW} signal in \eqref{equ:discrete_chirp} can be represented using \ac{SC-IFDM} structure after applying the inverse transform of \eqref{equ:simplifypp}. This yields the following expression
\begin{equation}
\begin{aligned}
    X_{\text{SC-IFDM}}^{\text{FMCW}}(k,l)&= \frac{1}{\sqrt{N}} \sum_{n=0}^{N-1}s^{\text{FMCW}}(l+n M)e^{-j2\pi\frac{k(l+nM)}{MN} }\\
    &=\frac{e^{j\pi (\frac{l^2}{MN}  ) }\omega_{k}^{l}}{\sqrt{N}} \sum_{n=0}^{N-1}e^{j2\pi\frac{\left(\frac{M}{2} n^2+(l-k) n\right)}{N} },
\label{equ:chirp_dft_1}
\end{aligned}
\end{equation} 
where $\omega_{k}^{l}= e^{j 2 \pi (\frac{-kl}{MN})}$. Note that \( e^{j\pi n} = 1 \) for even \( n \) and \( -1 \) for odd \( n \). Similarly, \( n^2 \) is even when \( n \) is even and odd when \( n \) is odd. Given that \( M \in \mathbb{Z} \), we find that \( e^{j\pi (M n^2)} = (-1)^{M n^2} = (-1)^{M n} \).
Incorporating this into \eqref{equ:chirp_dft_1}, we get
\begin{equation}
    X_{\text{SC-IFDM}}^{\text{FMCW}}(k,l)=\frac{e^{j\pi (\frac{l^2}{MN}  ) }\omega_{k}^{l}}{\sqrt{N}} \sum_{n=0}^{N-1}e^{j2\pi\frac{\left(\frac{M}{2} +l-k\right) n}{N} }.
\end{equation}
\hspace{1em}The elements of the matrix  $X_{\text{SC-IFDM}}^{\text{FMCW}}(k,l)$ are nonzero only when $\left[\frac{M}{2} +l-k\right]_N=0$ leading to a sparsity condition where the matrix is nonzero for certain values of $l$. Specifically, $X_{\text{SC-IFDM}}^{\text{FMCW}}(k,l)$ reduced to $s^{\text{FMCW}}(l)\omega_{k}^{l}$ for the valid solutions corresponding to each $l = 0, \dots, M-1$. This results in a sparse matrix representation, where only $M$ nonzero elements exist in the DFT-based space of the SC-IFDM.
Thus, FMCW signaling can be efficiently generated using SC-IFDM by selecting the first $M$ samples from $\mathbf{s}^{\text{FMCW}}$ according to this sparsity condition. This approach provides a computationally efficient way to combine the benefits of \ac{FMCW} and \ac{SC-IFDM} for \ac{JSAC} systems.Similarly, a down FMCW chirp is  represented using SC-IFDM structure as
\begin{equation}
   { X_{\text{SC-IFDM}}^{\text{FMCW}}} ^\prime(k,l)=\frac{e^{-j\pi (\frac{l^2}{MN}  ) }\omega_{k}^{l}}{\sqrt{N}} \sum_{n=0}^{N-1}e^{-j2\pi\frac{\left(\frac{M}{2} +l+k\right) n}{N} }.
\end{equation}

\subsection{Proposed JSAC Signal}
\begin{figure}[t!]
   \centering
\includegraphics[width=0.47\textwidth]{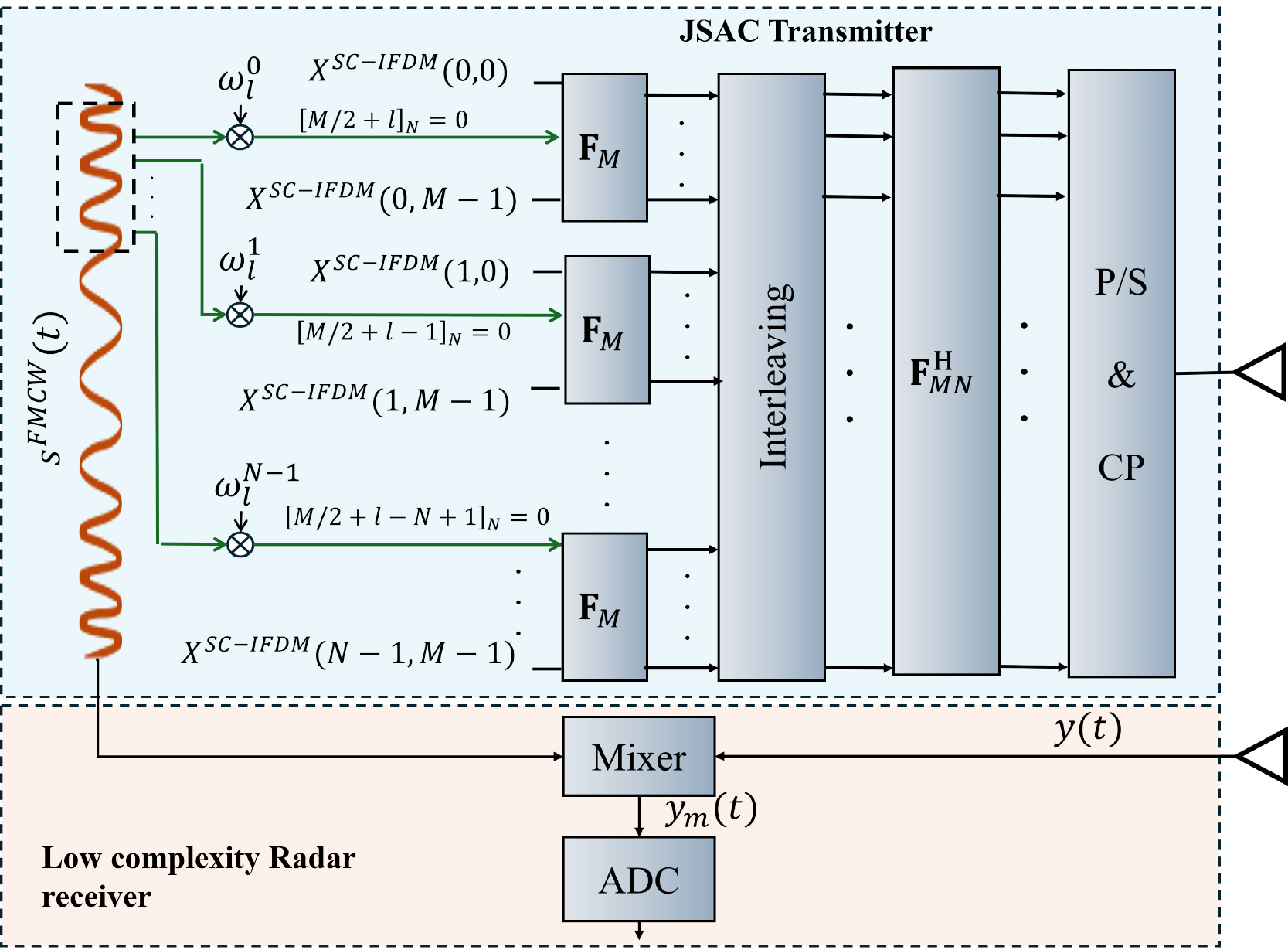}
    \caption{The proposed SC-IFDM-FMCW transceiver design. }
    \label{fig:transceiver}
\end{figure}

\hspace{1em}To generate the proposed \ac{JSAC} signal $X^{\text{comb}}(k,l)$, which integrates \ac{SC-IFDM} and \ac{FMCW} signals in an orthogonal manner, a specific allocation is employed. First, the \ac{SC-IFDM} signal is constructed by setting zeros at the indices where $\left[\frac{M}{2} +l-k\right]_N$, while the \ac{FMCW} signal is mapped to these zero positions. This ensures orthogonality between the two signals. The combined signal is defined as
\begin{equation}
    X^{\text{comb}}(k,l)= 
    \begin{cases} \sqrt{\boldsymbol{\psi}} s^{\text{FMCW}}(l)\omega_{k}^{l} 
    ,&\left[\frac{M}{2} +l-k\right]_N=0 \\X^{\text{SC-IFDM}}(k,l), & \text { otherwise },
    \end{cases}
\label{equ:SC-IFDM_CO_FMCW}
\end{equation}
where $\boldsymbol{\psi}$ is a design parameter representing the power assigned to \ac{FMCW} signal that will be used for sensing and also as pilots for channel estimation. 
Next, the combined signal $X^{\text{comb}}(k,l)$  is transformed into the time-domain signal $s^{\text{comb}}(l+nM)$ using the inverse transform from \eqref{equ:simplifypp}. Fig \ref{fig:transceiver} illustrates the implementation, \ac{FMCW} samples are allocated to the proper indices of each \ac{DFT} block of the \ac{SC-IFDM}.
Finally, the signal undergoes \ac{CP} insertion to ensure channel circularity. However, the potential discontinuity introduced by the \ac{CP} leads to discontinuities between the $i$-th and the $(i+1)$-th \ac{FMCW} signal, which decreases the sensing performance. Therefore, a proper time shift needs to be introduced to the $i$-th \ac{FMCW} in the \ac{DFT} domain to compensate for the \ac{CP} duration $L_{cp}$. The $i$-th \ac{FMCW} with the shift can be given as 
\begin{figure}[t!]
   \centering
\includegraphics[width=0.44\textwidth]{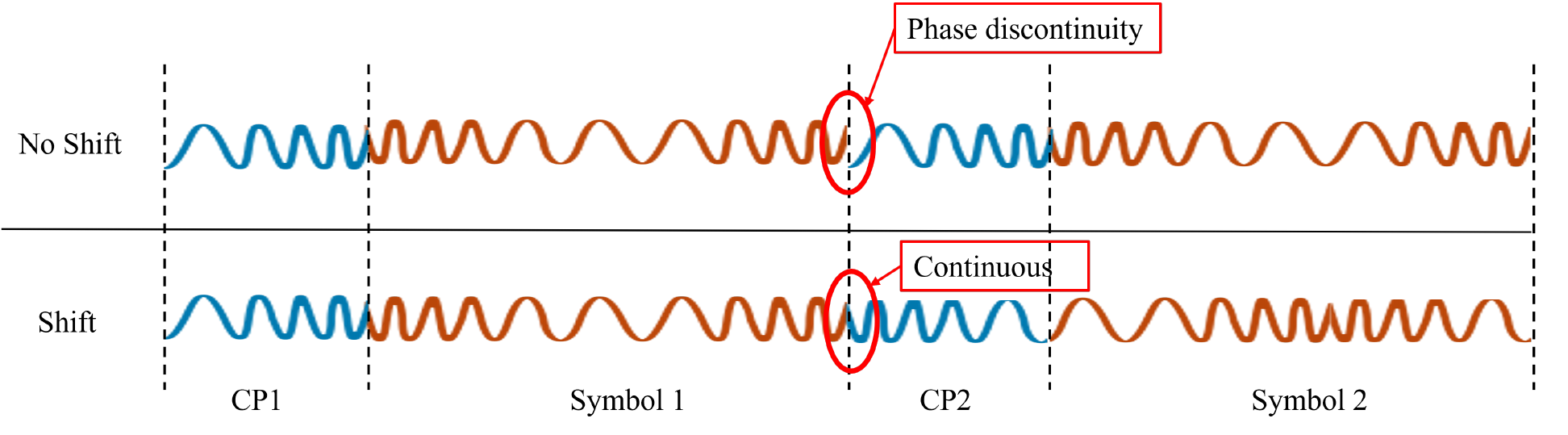}
    \caption{CP addition before and after phase shift. }
    \label{fig:CP}
\end{figure}
\begin{equation}
    X_{\text{SC-IFDM}}^{\text{FMCW},i}(k,l)= \frac{s^{\text{FMCW}}(l-i L_{cp})\omega_{k}^{l}}{\sqrt{N}} \sum_{n=0}^{N-1}e^{j2\pi\frac{\left(\frac{M}{2} +l-k-i L_{cp}\right) n}{N} }.
\label{equ:chirp_cp}
\end{equation}
\hspace{1em}Note that shifting the \ac{FMCW} signal in time by the \ac{CP} duration is a straightforward operation in \ac{SC-IFDM}. This is achieved by taking the $(l-iL_{cp})$ samples from the \ac{FMCW} signal and placing them in the indices following \( \left[\frac{M}{2} + l - k-iL_{cp}\right]_N = 0 \). Fig.\ref{fig:CP} illustrates how this time shift ensures that the \ac{FMCW} signal remains continuous after adding the \ac{CP}. As a result, no additional processing or \ac{CP} removal is required at the radar receiver side, allowing the signal to be directly mixed in the analog domain. This approach results in a low-complexity radar receiver design, as depicted in Fig. \ref{fig:transceiver}. 

\section{Channel}
\begin{figure*}[t!]
    \centering
    \subfigure[]{\includegraphics[width=0.245\textwidth]{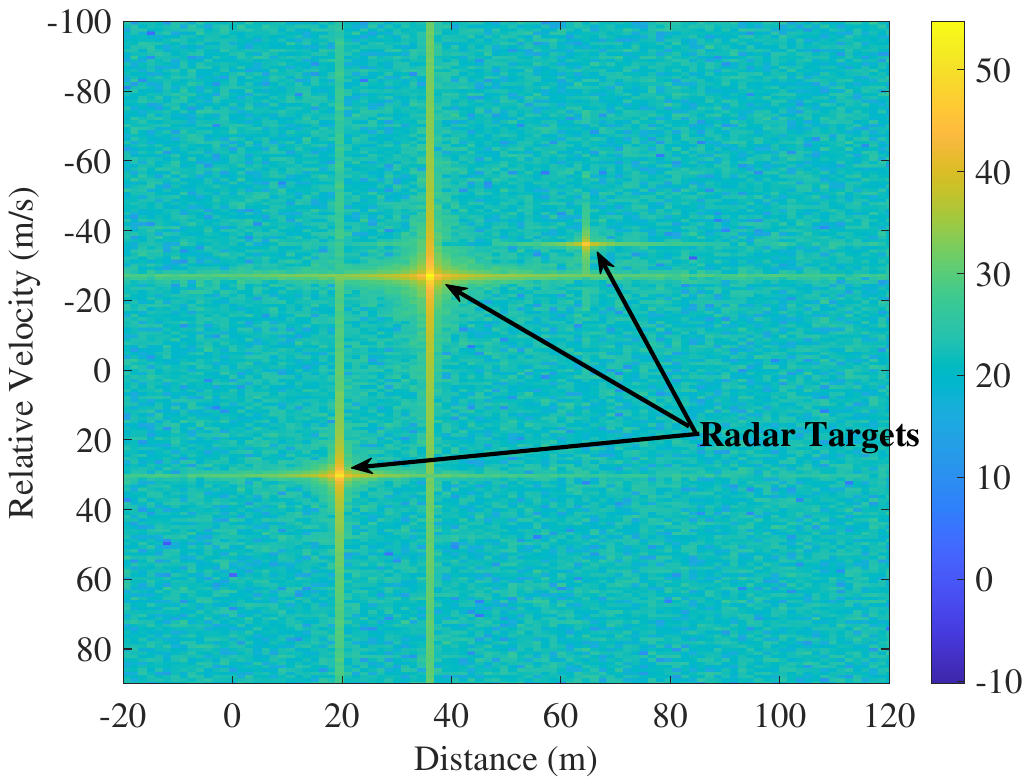}\label{fig:sub-first}}
   \subfigure[]{\includegraphics[width=0.245\textwidth]{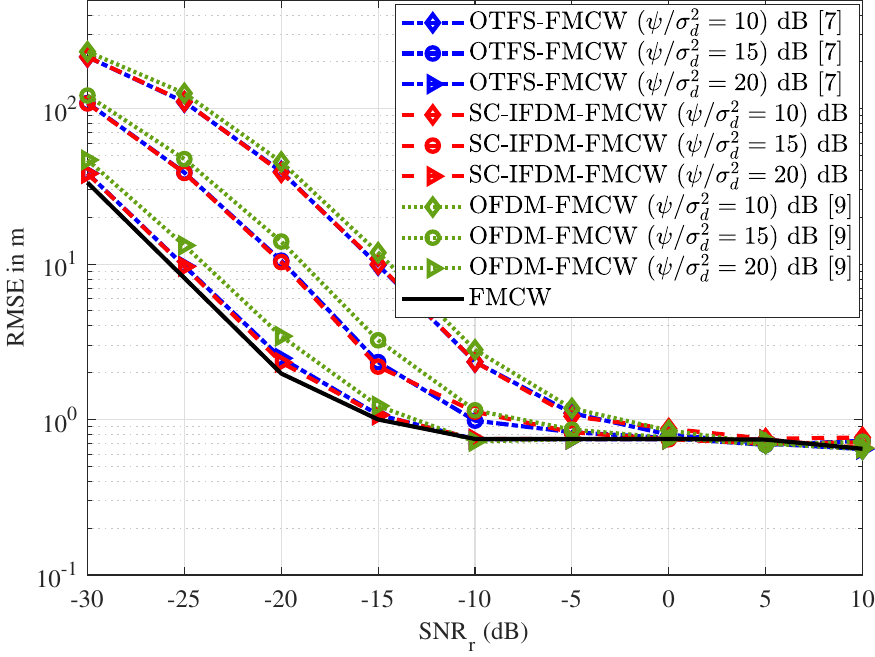}\label{fig:sub-second}} 
    \subfigure[]{\includegraphics[width=0.245\textwidth]{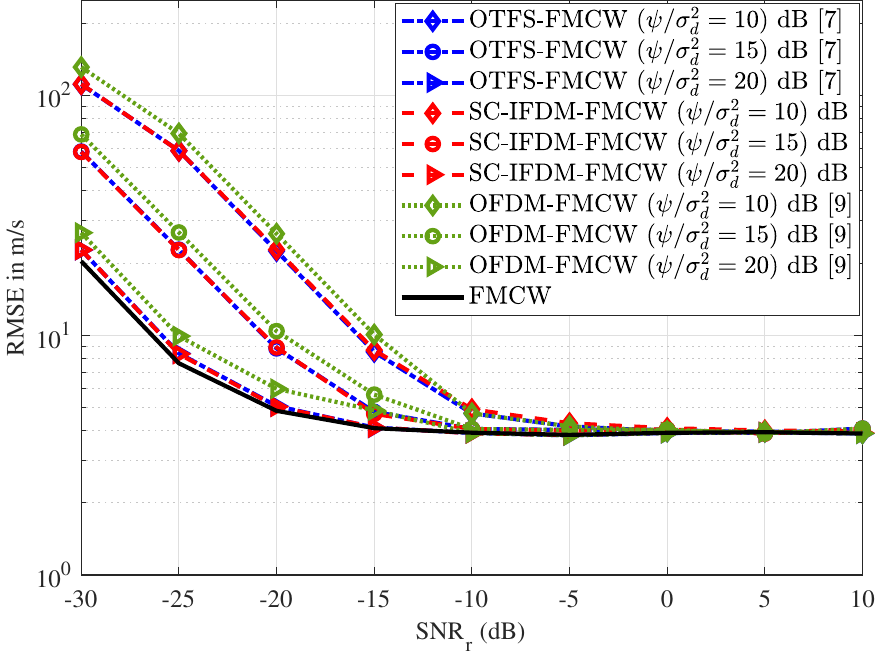}\label{fig:sub-third}}
     \subfigure[]{\includegraphics[width=0.245\textwidth]{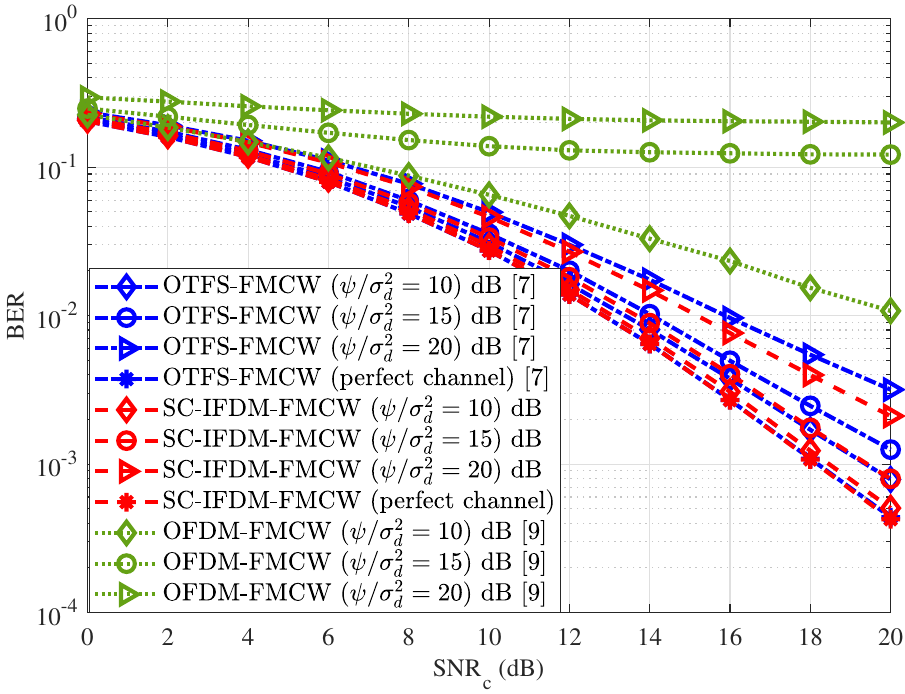}\label{fig:sub-fourth}}
    \caption{(a) Range Doppler map  (b) Range RMSE performance  (multiple targets) (c) Velocity RMSE performance  (multiple targets) (d) BER vs SNR.}
    \label{fig:sens_perf}
\end{figure*}
\subsection{FMCW Processing}
\hspace{1em}Consider a radar channel that models $Q$ targets with range $d_q$ and velocity $v_q$ as follows 
\begin{equation}
    h(\tau,\nu) = \sum_{q=0}^{Q-1}h_q\delta(t-\tau_q)e^{j2\pi\nu_q(t-\tau_q)},
    \label{equ:channel}
\end{equation} 
where $h_q$ is the complex channel gain corresponding to the $q$-th path, $\tau_q = 2d_q/c$ denotes the round-trip delay, $\nu_q = 2v_q f_c/c$ is the Doppler shift, $f_c$ is the carrier frequency, and $c$ is the speed of light. The radar channel's output is then expressed as
\begin{equation}
    y(t) = \sum_{q=0}^{Q-1}h_q s^{comb}(t-\tau_q)e^{j2\pi\nu_q(t-\tau_q)}.
    \label{equ:channel_out}
\end{equation}
\hspace{1em}To extract the beat frequency $f_B$ and the Doppler shift $f_D$, which carries the information about the time delay and Doppler shift induced by the targets, the received reflected signal is mixed with the conjugate of the transmitted chirp, $\text{conj}(s^{\text{FMCW}})$. For an up-chirp, the output is 
\begin{equation}
    y_m(t) =\sum_{q=0}^{Q-1} \tilde{h}_q e^{j2\pi(-\eta\tau_q + \nu_q)t} + I(t),
\label{equ:15}
\end{equation}
where $\tilde{h}_q= h_q e^{j\pi \eta \tau^2} e^{-j2\pi v_q \tau_q}$ is the complex amplitude of the $q$-th path, $\eta$ is the chirp slope, and $I(t)$ is the interference  from \ac{SC-IFDM} which can be given as
\begin{equation}
    I(t) = w(t)+\sum_{q=0}^{Q-1} h_q s^{\text{SC-IFDM}}(t - \tau_q) e^{j2\pi\nu_q(t-\tau_q)} \cdot e^{-j\pi  \eta t^2 }.
\end{equation}
\hspace{1em}The radar data obtained from the up-chirp, as outlined in \cite{kim2013simulation}, is expressed as \( f_{bu} = f_B - f_D \). Similarly, processing the down-chirp results in \( f_{bd} = f_B + f_D \). These frequencies are used to calculate the range and velocity. 
\subsection{Communication channel}
\hspace{1em}
Consider a doubly selective \( R \)-tap channel, which can be expressed in discrete time domain as
\begin{equation}
h(p) = \sum_{r=0}^{R-1} h_r e^{j2\pi \frac{k_r (p-l_r)}{MN}} \delta(p-l_r),    
\end{equation}
where \( h_r \) denote the complex channel coefficient, while \( l_r \) and \( k_r \) represent the integer delay and Doppler shifts introduced by the channel, respectively. The received signal is given as

\begin{equation}
\begin{aligned}
   Y^{\text{comb}}(k,l)&= \sum_{r=0}^{R-1} h_r e^{j2\pi \frac{k_r}{N}\frac{l-l_r}{M}}\Lambda_r(k,l)\\ &\times X^{\text{comb}} ([k-k_r]_N,[l-l_r]_M)+W(k,l),
   \label{equ:y_com_c}
   \end{aligned}
\end{equation}
where $\Lambda_r(k, l)$ is defined as $1$ for $l_r\leq l \leq M$ and as $e^{-j2\pi \frac{k}{N}}$ for $ 0 \leq l \leq l_r$, and \textcolor{blue}{$W(k,l)$} is additive white Gaussian noise of zero mean and variance $\sigma^2$.
It is demonstrated that any delay or Doppler shift introduces a shift in both the data and chirp components. Since channel estimation relies on the chirp in the DFT domain, this causes interference between the data and chirp, ultimately degrading the accuracy of the channel estimation. The interfered chirp elements in the received signal, \( Y_{dp} \), can be extracted by satisfying the condition \([\frac{M}{2} + l + k_r - l_r - k]_N = 0\), leading to the following result given as
\begin{equation}
\begin{aligned}
    Y_{dp}(k, l)=Y^{\text{FMCW}}(k,l)+\underbrace{Y^{\text{SC-IFDM}}(k,l)}_{\text{data to pilot interference}} + \underbrace{W(k,l)}_{\text{noise}}, 
\end{aligned}
\end{equation}
with the received \ac{FMCW} signal  written as
\begin{equation}
    Y^{\text{FMCW}}(k,l)=\sum_{r=0}^{R-1} \sqrt{\boldsymbol{\psi}}\bar{h_r} e^{j 2\pi  \frac{l(k_r-l_r)+l_r^2 -k_r l_r}{MN}}  e^{j 2\pi \frac{l^2}{MN}}
\end{equation}
where $\bar{h_r}=h_r \Lambda_r(k-k_r,l-l_r)$. It is observed that the channel shifts the chirp location by \( k_r - l_r \), causing channel estimation ambiguity, similar to the issue described in \ac{OCDM} and discussed in \cite{haif2024novel}. 
\subsection{Proposed Channel Estimation}
\hspace{1em}Due to the behavior of chirps in doubly-selective channels, a novel channel estimation approach is needed. To separate  data from the received \ac{FMCW} pilots, we introduce a transform that extracts every possible chirp in each $M$ samples as follows 
\begin{equation}
\hat{Y}_1(\beta_1,\alpha_1)= \frac{1}{M}\sum_{l=0}^{M-1} Y_{dp}(k,l)
 ~ e^{-j \pi \frac{(l-\alpha_{1} N - \beta_1)}{MN}},
 \label{equ:transform}
\end{equation}
with $
\frac{M}{2} + l + k_r - l_r - k = \alpha_1 N + \beta_1, $ where $\alpha_1$ and $\beta_1$ are integers satisfying $\alpha_1 = 0, \ldots, M-1$ and $\beta_1 = 0, \ldots, N-1$. For each $\beta_1$, if $l_r-k_r=\alpha_1 N +\beta_1$, the sum becomes  
\begin{equation}
\hat{Y}_1(\beta_1,\alpha_1)= \sum_{r=0}^{R-1}\bar{h_r} e^{j \pi\frac{-k_r^2}{MN}}+\hat{Y}^{\text{SC-IFDM}}(\beta_1,\alpha_1)+
    \hat{W}(\beta_1,\alpha_1),
\end{equation}
where $\hat{W}(\beta_1,\alpha_1)$ is the transformed noise. Since the data consist of uniformly distributed QAM symbols with power $\sigma_d^2$ and zero mean, the expectation 
$\mathbf{E}\left[\sum_{r=0}^{R-1} Y^{\text{SC-IFDM}}(\beta_1,\alpha_1)
~ e^{-j \pi \frac{(l-\alpha_1 N - \beta_1)}{MN}}\right]$ is zero.  The variance of $\hat{Y}_1$ is given as 
\begin{equation}
\text{var}\{ \hat{Y}_1\}=\frac{\sigma_d^2 (\sum_{r=0}^{R-1}h_r^2) + \sigma^2 }{ \boldsymbol{\psi} M}.
\end{equation}
\hspace{1em}The estimated value $\hat{Y}_1(\beta_1,\alpha_1)$ is approximately $\bar{h_r} e^{j \pi\frac{-k_r^2}{MN}}$ for $l_r \neq k_r$ and  $\sum_{r=0}^{R-1}\bar{h_r} e^{j \pi\frac{-k_r^2}{MN}}$ for $l_r = k_r$, which is not sufficient for estimating $l_r, k_r$ and $h_r$. Therefore, in this particular case we consider an extra step to estimate the delay and Doppler shifts where two SC-IFDM symbols
are used, one will have an up-chirp and the other a down-chirp.  Similarly, a received down-chirp can be given as 
\begin{equation}
   {Y^{\text{FMCW}}} ^\prime(k,l)=\sum_{r=0}^{R-1} \sqrt{\boldsymbol{\psi}}\bar{h_r} e^{j 2\pi  \frac{l(k_r+l_r)-l_r^2-k_r l_r}{MN}} e^{j 2\pi \frac{-l^2}{MN}},
\end{equation} 
which also faces an ambiguity in differentiating delay and Doppler shifts in the case of \( l_r = -k_r \). However, using the same transform as in \eqref{equ:transform}, we can derive the following
\begin{equation}
    \begin{cases}
        k_r-l_r=\alpha_1 N + \beta_1\\
        k_r+l_r=\alpha_2 N + \beta_2
    \end{cases},
\end{equation}
where $\alpha_1, \beta_1$ and $\alpha_2, \beta_2$ are the estimated shifts from the chirp in symbol 1 and symbol 2, respectively. For each $\alpha_1, \beta_1$, we check if a solution for $\alpha_2, \beta_2$ exists such that the delay and Doppler shifts are integers. Once these shifts are determined, the overlapping channel coefficients can be estimated using a symbol that is not affected by overlap.

\section{SIMULATIONS RESULTS}
\hspace{1em}In this section, we evaluate the communication and sensing performance of the proposed waveform design. The simulation parameters are configured as follows: a carrier frequency \( f_c \) of \( 77 \, \mathrm{GHz} \), chosen for its high range-Doppler resolution but not restricted to this value, a bandwidth of \( 200 \, \mathrm{MHz} \), and each symbol using \( 32 \times 32 \) samples. The simulation uses \(100\) symbols and benchmarks performance against \cite{zegrar2022otfs}, \cite{zegrar2024novel}, and \cite{bouziane2024novel} for both communication and sensing, including a comparison with traditional FMCW for sensing.

For the sensing evaluation, we present the range-velocity plots. The range-velocity map, as shown in Fig. \ref{fig:sub-first}, is extracted using the conventional radar processing for the proposed JSAC scenario \cite{kim2013simulation}. The simulation targets were placed at distances uniformly distributed between 10 and 80 meters, with velocities ranging between -70 and 70 meters per second.

Fig. \ref{fig:sub-second} and \ref{fig:sub-third} compare the \ac{RMSE} for range and velocity between the proposed SC-IFDM-FMCW,  \cite{zegrar2024novel}, \cite{bouziane2024novel}, and traditional FMCW. The proposed JSAC waveform approaches \ac{FMCW} performance as the power ratio \( \frac{\boldsymbol{\psi}}{\sigma_d^2} \) increases especially at low radar \ac{SNR} (i.e. $\text{SNR}_r <0 $ dB) , demonstrating robustness across different sensing environments. It performs similarly to OTFS-FMCW due to its sparse and orthogonal chirp representation. SC-IFDM-FMCW outperforms OFDM-FMCW for both range and velocity, as OFDM-FMCW lacks orthogonality between data and chirp, leading to interference.\\
We assess the \ac{BER} versus the communication \ac{SNR}$_c$ performance of the proposed JSAC waveform and compare it with the approaches in \cite{zegrar2024novel} and \cite{bouziane2024novel}. Fig. \ref{fig:sub-fourth} presents \ac{BER} results for various power ratios (\( \frac{\boldsymbol{\psi}}{\sigma_d^2} \in \{10, 15, 20\} \) dB), considering the proposed channel estimation and perfect channel knowledge for OTFS-FMCW and SC-IFDM-FMCW.
\hspace{1em}The proposed SC-IFDM-FMCW achieves comparable \ac{BER} to OTFS-FMCW under perfect channel knowledge and surpasses it when using the proposed channel estimation. This improvement mitigates the effects of channel aging, as the radar-based delay and Doppler estimation method in \cite{zegrar2024novel} requires numerous symbols to accurately estimate shifts, particularly Doppler, which depends on phase changes. Additionally, SC-IFDM-FMCW outperforms OFDM-FMCW due to its lack of orthogonality and very poor performance in mobile environments. However, as \( \frac{\boldsymbol{\psi}}{\sigma_d^2} \) increases, \ac{BER} degrades due to higher FMCW interference, emphasizing the critical trade-off between sensing power and communication quality for system optimization.

 \begin{figure}[t]
   \centering
\includegraphics[width=0.35\textwidth]{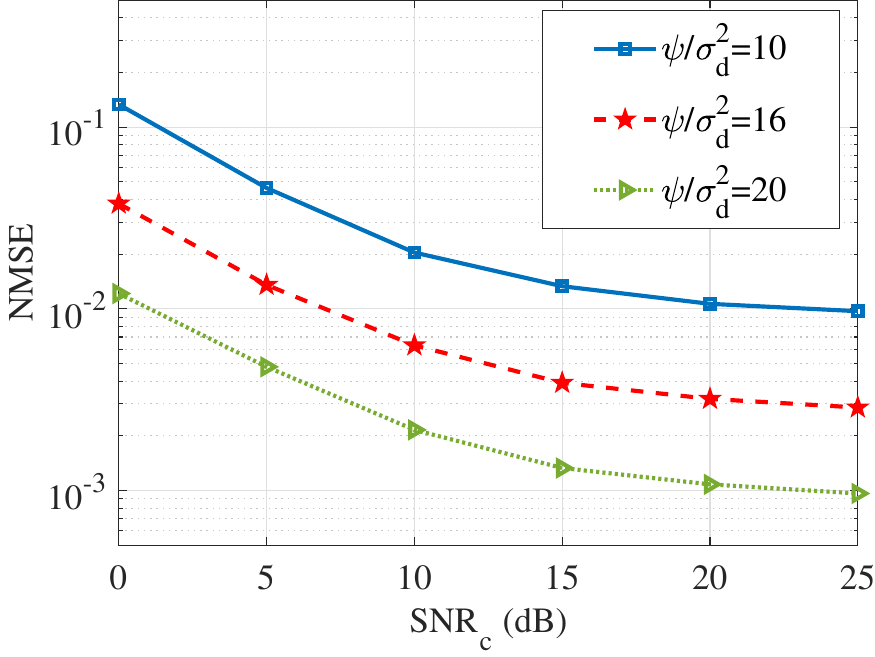}
    \caption{  Channel NMSE performance of the proposed SC-IFDM-FMCW estimation. }
    \label{fig:NMSE}
\end{figure}
Fig. \ref{fig:NMSE} illustrates the \ac{NMSE} performance of the proposed channel estimation technique across varying \ac{FMCW} and data power levels  \( \frac{\boldsymbol{\psi}}{\sigma_d^2} \). The results show that \ac{NMSE} improves as the pilot power ratio \( \frac{\boldsymbol{\psi}}{\sigma_d^2} \) increases, reflecting more precise channel estimation. This trend is consistent with the \ac{BER} performance observed in comparison to the method in \cite{zegrar2022otfs}. However, the overall \ac{BER} performance deteriorates due to increased interference power, underscoring the importance of careful optimization to balance sensing and communication efficiency. 
\section{Conclusion}
\hspace{1em}This paper presents a novel \ac{JSAC} waveform, \ac{SC-IFDM}-\ac{FMCW}, designed to support both communication and sensing. The approach combines \ac{SC-IFDM} for high data rates and robust mobile performance with \ac{FMCW} chirp signals for high-resolution sensing, enabling seamless coexistence of both functionalities. The design ensures backward compatibility with existing radar and communication systems, facilitating integration into radar receivers without added complexity. It also addresses challenges like channel aging and interference through an enhanced channel estimation technique, ensuring reliable performance in dynamic environments. Our results show that \ac{SC-IFDM}-\ac{FMCW} outperforms OFDM-FMCW in sensing accuracy and surpasses both OFDM-FMCW and OTFS-FMCW in \ac{BER} performance. The power ratio \( \frac{\boldsymbol{\psi}}{\sigma_d^2} \) is identified as a critical factor in balancing sensing and communication, highlighting the need for optimization. Future work will optimize waveform parameters, including power ratio, using AI to enhance system performance in real-world conditions. This research positions \ac{SC-IFDM}-\ac{FMCW} as a key technology for advancing communication and sensing integration in 6G.



\end{document}

%% file: acro.tex
\DeclareAcronym{6G}{short = 6G,long  = sixth Generation ,tag    = abbrev}
\DeclareAcronym{JSAC}{  short = JSAC,  long  = joint sensing and communication   ,  tag    = abbrev}
\DeclareAcronym{OTFS}{  short = OTFS,  long  = orthogonal time-frequency space ,  tag    = abbrev}
\DeclareAcronym{OFDM}{short = OFDM,long  = orthogonal frequency division multiplexing ,tag    = abbrev}
\DeclareAcronym{FMCW}{short = FMCW,long  = frequency modulated continuous wave ,tag    = abbrev}
\DeclareAcronym{SNR}{short = SNR,long  = signal-to-noise ratio ,tag    = abbrev}
\DeclareAcronym{SINR}{short = SINR,long  = signal-to-interference-and-noise ratio ,tag    = abbrev}
\DeclareAcronym{PAPR}{short = PAPR,long  = peak to average power ratio ,tag    = abbrev}
\DeclareAcronym{DFT}{short = DFT,long  = discrete Fourier transform ,tag    = abbrev}
\DeclareAcronym{CP}{short = CP,long  = cyclic prefix , tag  = abbrev}
\DeclareAcronym{IDFT}{short = IDFT,long  = inverse discrete Fourier transform ,tag    = abbrev}
\DeclareAcronym{FFT}{short = FFT,long  = fast Fourier transform ,tag    = abbrev}
\DeclareAcronym{IFFT}{short = IFFT,long  = inverse fast Fourier transform ,tag    = abbrev}
\DeclareAcronym{NMSE}{short = NMSE,long  = normalized mean square error ,tag    = abbrev}
\DeclareAcronym{AWGN}{short = AWGN,long  = additive white Gaussian noise,tag    = abbrev}
\DeclareAcronym{OCDM}{  short = OCDM,  long  = orthogonal chirp division multiplexing ,  tag    = abbrev}
\DeclareAcronym{BER}{  short = BER,  long  = bit error rate ,  tag    = abbrev}
\DeclareAcronym{RMSE}{  short = RMSE,  long  = root mean
squared error ,  tag    = abbrev}
\DeclareAcronym{SC-IFDM}{short = SC-IFDM,long  = single-carrier interleaved frequency division multiplexing,tag    = abbrev}